\documentclass[twocolumn, aps, prl, showpacs, floatfix]{revtex4}

\usepackage{multirow}
\usepackage{epsfig}
\usepackage{amsmath}

\usepackage{subfigure}
\usepackage{latexsym}
\usepackage{feynmp}

\usepackage[normalem]{ulem}  
\usepackage[dvips]{color} 

\renewcommand\sout{\bgroup \color{red} \ULdepth=-.5ex \ULset}

\begin{document}

\title{Enhanced production of $\psi(2S)$ mesons in heavy ion collisions}

\author{Sungtae Cho}
\affiliation{Institute of Physics and Applied Physics, Yonsei
University, Seoul 120-749, Korea}

\begin{abstract}

We study the production of a $\psi(2S)$ meson in heavy ion
collisions. We evaluate Wigner functions for the $\psi(2S)$ meson
using both Gaussian and Coulomb wave functions, and investigate
the wave function dependence in the $\psi(2S)$ meson production by
recombination of charm and anti-charm quarks. The enhanced
transverse momentum distribution of $\psi(2S)$ mesons compared to
that of $J/\psi$ mesons, originated from wave function
distributions of the $\psi(2S)$ and $J/\psi$ meson in momentum
space, provides a plausible explanation for the recent measurement
of the nuclear modification factor ratio between the $\psi(2S)$
and $J/\psi$ meson.

\end{abstract}



\maketitle

Since $J/\psi$ suppression was proposed as a possible signature
for the formation of a system of deconfined quarks and gluons, the
so called quark-gluon plasma (QGP) \cite{Matsui:1986dk} $J/\psi$
has been one of standard probes in understanding many aspects of
heavy ion collision experiments. Moreover, its excited states as
well have become important probes of the QGP since they are
expected to melt sequentially in the medium due to their different
binding energies, and therefore they can play important roles in
investigating the properties of the QGP \cite{Satz:2005hx,
Karsch:2005nk, Mocsy:2007jz}.

One of the important measures for the effects of the QGP on the
$J/\psi$ production is a nuclear modification factor $R_{AA}$,
defined as the ratio of the yield of a hadron in heavy ion
collisions to that in p+p collisions scaled by the number of
binary collisions. Measurements of the $J/\psi$ $R_{AA}$ at the
Relativistic Heavy Ion Collider (RHIC) show that the $J/\psi$
production is suppressed significantly in heavy ion collisions,
and the $R_{AA}^{J/\psi}$ decreases with the increasing centrality
\cite{Adare:2006ns}. Although we still observe the $J/\psi$
suppression at higher energies, we see the noticeable difference
in the $R_{AA}^{J/\psi}$ dependence on the centrality; the
$R_{AA}$ for $J/\psi$ mesons measured at the Large Hadron Collider
(LHC) is independent of the collision centrality at forward
rapidity \cite{Abelev:2012rv, Abelev:2013ila}.

The less suppression of $J/\psi$ mesons at LHC supports
possibilities of $J/\psi$ production from charm quarks in the QGP
\cite{Thews:2000rj, BraunMunzinger:2000px, Andronic:2007bi}. The
larger charm quark density at LHC compared to that at RHIC results
in enhanced chances for the formation of $J/\psi$ from the QGP.
Moreover, the strong $J/\psi$ suppression at high transverse
momentum $p_T$ measured by CMS \cite{Chatrchyan:2012np} together
with the $R_{AA}^{J/\psi}$ at low $p_T$ measured by ALICE
\cite{Abelev:2013ila} and the measurement of $J/\psi$ elliptic
flows at LHC \cite{ALICE:2013xna} provide solid evidences for the
$J/\psi$ production from charm quarks by recombination.

In addition, the measurement of the $\psi(2S)$ $R_{AA}$ relative
to the $R_{AA}^{J/\psi}$ at LHC provides chances to investigate
the production of charmonium states from the QGP. The less
suppression of the $\psi(2S)$ than in p+p collisions compared to
the $J/\psi$ with the increasing centrality recently measured by
CMS \cite{CMS} has not been clearly understood. In this work we
argue that the increased double ratio
$R_{AA}^{\psi(2S)}/R_{AA}^{J/\psi}$ at central collisions is also
due to their production from charm quarks by recombination.

The recombination picture describes the formation of hadrons as
the process of coalescing constituent quarks in the QGP into
hadrons using phase space density functions of the constituents
and of the produced hadron, or the Wigner distribution functions
composed of the overlap between wave functions. Attentions paid to
the production of hadrons from constituent quarks at low and
intermediate $p_T$ have leaded to the explanation of the quark
number scaling of elliptic flows of identified hadrons
\cite{Molnar:2003ff} as well as the enhanced production of baryons
at midrapidity \cite{Greco:2003xt, Greco:2003mm, Fries:2003vb,
Fries:2003kq}.

In the recombination process, on the other hand, when different
hadrons are produced from same constituents, hadron Wigner
distributions become important, and play central roles in
describing the properties of produced hadrons. The production of
charmonium states from same charm quark distributions in the QGP
by recombination is such case. We study here the hadron wave
function dependence in charmonium production while we investigate
the production of $\psi(2S)$ mesons.

As simple forms of the wave function, the Gaussian-type functions
have been used in many literatures. In particular, the
Gaussian-type Wigner function made up of simple harmonic
oscillator wave functions reflects the hadron size through the
oscillator frequency of the wave function $\omega$, and takes into
account an orbital excitation of the produced hadron by
introducing $s$-, $p$- \cite{Baltz:1995tv}, and $d$-wave
\cite{Cho:2011ew} harmonic oscillator wave functions,
\begin{eqnarray}
&& W_s(\vec r, \vec k) = 8 e^{-\frac{r^2}{\sigma^2}-k^2
\sigma^2} \nonumber \\
&& W_p(\vec r, \vec k) = \bigg( \frac{16}{3} \frac{r^2}{\sigma^2}
-8+\frac{16}{3} \sigma^2 k^2 \bigg)
e^{-\frac{r^2}{\sigma^2}-k^2 \sigma^2} \nonumber \\
&& W_d(\vec r,\vec k) =
\frac{8}{15}\Big(4\frac{r^4}{\sigma^4}-20\frac{r^2}{\sigma^2}
+15-20\sigma^2k^2+4\sigma^4k^4 \nonumber \\
&& \qquad\quad\quad+16r^2k^2-8(\vec r\cdot\vec k)^2\Big)
e^{-\frac{r^2}{\sigma^2}-k^2\sigma^2} , \label{WigGau}
\end{eqnarray}
where $\sigma^2=1/(\mu\omega)$ with the reduced mass $\mu$.

We apply here the above Gaussian Wigner function to describe the
formation of $J/\psi$ and $\chi_c$ mesons. Furthermore, we
construct also Wigner functions based on Coulomb wave functions
since we expect that the wave function of the charmonium states
formed through a color cental potential between charm and
anti-charm quarks is closer to the color Coulomb wave function
governed by a chromo-electric field rather than the Gaussian wave
function based on the simple harmonic oscillator potential. We
investigate the formation of charmonium states based on both
Gaussian and Coulomb Wigner functions, and explain the enhanced
production of the loosely bound charmonium state, the $\psi(2S)$
compared to the $J/\psi$ at the most central heavy ion collisions.

Wigner functions based on Coulomb wave functions have been unclear
until recently despite the well-known analytic form of Coulomb
wave functions. The systematic way of generating Wigner functions
for arbitrary states of hydrogen atom has been presented in 2006
\cite{Praxmeyer:2006}. The key idea is to evaluate them in
momentum space using the Feynman parametrization. Two joint
Coulomb wave functions in momentum space resemble the
multiplication of two propagators often met in field theory.

We sketch below the method introduced in Ref.
\cite{Praxmeyer:2006}, and present explicit $1s$ and $2s$ Wigner
functions constructed from Coulomb wave functions. We begin with
the Wigner function defined as
\begin{eqnarray}
&& W_{\psi}(\vec r, \vec k)=\int d^3\vec q \psi^*(\vec r+\vec
q/2)e^{i\vec k\cdot\vec q}\psi(\vec r-\vec q/2) \nonumber \\
&& \qquad\qquad =\int \frac{d^3\vec q}{(2\pi)^3}
\tilde{\psi}^*(\vec k+\vec q/2)e^{-i\vec r\cdot\vec
q}\tilde{\psi}(\vec k-\vec q/2), \quad \label{wigner}
\end{eqnarray}
which has been normalized to satisfy the condition $\int
W_{\psi}(\vec r, \vec k)d^3\vec r d^3\vec k=(2\pi)^3$. After
plugging the wave function of the ground state in momentum
representation $\tilde{\psi}_{1S}(\vec
k)=8\sqrt{\pi}a_0^{3/2}/(1+k^2a_0^2)^2$ into Eq. (\ref{wigner}) we
get
\begin{eqnarray}
&& W_{\psi_{1S}}(\vec r, \vec k) \nonumber \\
&& =\frac{64}{\pi^2a_0^5}\int d^3\vec q' \frac{e^{-2i\vec
r\cdot(\vec q'-\vec k)}}{(1/a_0^2+\vec {q'}^2)^2(1/a_0^2+(\vec
q'-2\vec k)^2)^2}.
\end{eqnarray}
With the help of the Feynman parametrization,
\begin{equation}
\frac{1}{A^{\alpha}B^{\beta}}=\frac{\Gamma(\alpha+\beta)}{\Gamma(
\alpha)\Gamma(\beta)}\int_0^1du\frac{u^{\alpha-1}(1-u)^{\beta-1}}
{(Au+B(1-u))^{\alpha+\beta}}, \label{feynman}
\end{equation}
we disentangle the denominator and obtain
\begin{eqnarray}
&& W_{\psi_{1S}}(\vec r, \vec k)=\frac{64}{\pi^2a_0^5} \Gamma(4)
\int_0^1 du u(1-u) e^{-2i(1-2u)\vec r\cdot\vec k} \nonumber \\
&& \qquad\qquad\quad \times \int d^3\vec s \frac{e^{-2i\vec
r\cdot\vec s}}{(s^2+1/a_0^2+4u(1-u)k^2)^4},
\end{eqnarray}
where the change of a variable from $\vec q'$ to $\vec s=\vec
q'-2(1-u)\vec k$ has been made. We carry out the integration over
$\vec s$ analytically, and finally get
\begin{eqnarray}
&& W_{\psi_{1S}}(\vec r,\vec k)=\frac{16}{a_0^5}\int_0^1du u(1-u)
e^{-2i(1-2u)\vec k\cdot\vec r}e^{-2rC(u)} \nonumber \\
&& \qquad\qquad\quad \times
\Big(\frac{3}{C(u)^5}+\frac{6}{C(u)^4}r+\frac{4}{C(u)^3} r^2\Big),
\label{wigner1s}
\end{eqnarray}
with $C(u)=(1/a_0^2+4u(1-u)k^2)^{1/2}$. Similarly, with the second
excited state wave function in momentum representation,
$\tilde{\psi}_{2S}(\vec k)=\sqrt{\pi}(2a_0)^{3/2}
(k^2a_0^2-1/4)/(k^2a_0^2+1/4)^3$ we obtain
\begin{widetext}
\begin{eqnarray}
&& W_{\psi_{2S}}(\vec r,\vec k)=\frac{1}{32a_0^9}\int_0^1du
u^2(1-u)^2e^{-2i(1-2u)\vec k\cdot\vec
r}\Big(\frac{105}{D(u)^9}+\frac{210}{D(u)^8}r+\frac{180}{D(u)^7}r^2+
\frac{80}{D(u)^6}r^3+\frac{16}{D(u)^5}r^4\Big)e^{-2rD(u)} \nonumber \\
&& \qquad\qquad-\frac{1}{4a_0^7}\int_0^1du u(1-u)e^{-2i(1-2u)\vec
k\cdot\vec
r}\Big(\frac{15}{D(u)^7}+\frac{30}{D(u)^6}r+\frac{24}{D(u)^5}r^2+
\frac{8}{D(u)^4}r^3\Big)e^{-2rD(u)} \nonumber \\
&& \qquad\qquad+\frac{2}{a_0^5}\int_0^1du u(1-u)e^{-2i(1-2u)\vec
k\cdot\vec
r}\Big(\frac{3}{D(u)^5}+\frac{6}{D(u)^4}r+\frac{4}{D(u)^3}r^2\Big)
e^{-2rD(u)}, \label{wigner2s}
\end{eqnarray}
\end{widetext}
with $D(u)=(1/(2a_0)^2+4u(1-u)k^2)^{1/2}$.

We also construct the $2s$ state Wigner function using a Gaussian
wave function. The proper wave function would be
$\psi_{10}=\sqrt{2/3}(1/\pi\sigma^2)^{3/4}e^{-r^2/2\sigma^2}(-r^2/\sigma^2+3/2)$,
where the subscript $10$ in $\psi$ represents the quantum number
$kl$ in the 3-dimensional harmonic oscillator satisfying
$E_n=(n+3/2)\hbar\omega=(2k+l+3/2)\hbar\omega$. $\psi_{10}$ is the
first excited state of the ground state $\psi_{00}$ with the
lowest angular momentum, and therefore can play a role of the wave
function for the $\psi(2S)$ meson. From the definition of the
Wigner function, Eq. (\ref{wigner}), we get {\allowdisplaybreaks
\begin{eqnarray}
&& W_{\psi_{10}}(\vec r,\vec k) =
\frac{16}{3}\Big(\frac{r^4}{\sigma^4}-2\frac{r^2}{\sigma^2}
+\frac{3}{2}-2\sigma^2k^2+\sigma^4k^4 \nonumber \\
&& \qquad\qquad\quad-2r^2k^2+4(\vec r\cdot\vec k)^2\Big)
e^{-\frac{r^2}{\sigma^2}-k^2\sigma^2} \label{wigner10}.
\end{eqnarray}}
The Wigner function for $\psi_{10}$, Eq. (\ref{wigner10}) is very
similar in form to $W_d(\vec r, \vec k)$, Eq. (\ref{WigGau}). This
must be due to degenerate energy states between the $d$-wave state
($k$=0, $l$=2) and the first excited state of the $s$-wave state
($k$=1, $l$=0) in the 3-dimensional harmonic oscillator.

Using these Wigner functions we evaluate the transverse mometum
distribution of charmonia, $J/\psi$, $\chi_c$, and $\psi(2S)$
mesons produced from charm and anti-charm quarks by recombination.
We start with the basic equation in the coalescence model
\cite{Greco:2003mm},
\begin{eqnarray}
&& N_\psi=g_\psi\int p_c\cdot d\sigma_c p_{\bar{c}}\cdot
d\sigma_{\bar{c}} \frac{d^3\vec p_c}{(2\pi)^3E_c}\frac{d^3\vec
p_{\bar{c}}}{(2\pi)^3E_{\bar{c}}} \nonumber \\
&& \qquad \times f_c(r_c, p_c)f_{\bar{c}}(r_{\bar{c}},
p_{\bar{c}})W_\psi(r_c, r_{\bar{c}} ; p_c, p_{\bar{c}}),
\label{CoalGen}
\end{eqnarray}
with space-like hypersurface elements $d\sigma$ and covariant
distribution functions for a(n) (anti-)charm quark
$f_{c(\bar{c})}(r, p)$ satisfying the normalization condition
$\int p\cdot d\sigma d^3\vec p/((2\pi)^3E)f_{c(\bar{c})}(r,
p)=N_{c(\bar{c})}$, the number of (anti-)charm quarks available in
the system. The statistical factor $g_\psi$ accounts for the
possibility of forming a charmonium from constituent quarks, e.g.,
$g_{J/\psi}=3/36$. When the non-relativistic limit is taken, the
above equation is reduced to \cite{Greco:2003mm, Greco:2003xt,
Oh:2009zj}
\begin{eqnarray}
&& \frac{d^2N_\psi}{d^2\vec p_T}=\frac{g_\psi}{V}\int d^3\vec r
d^2\vec p_{cT}d^2\vec p_{\bar{c}T}\delta^{(2)}(\vec p_T-\vec
p_{cT}-\vec p_{\bar{c}T}) \nonumber \\
&& \qquad\quad \times \frac{d^2N_c}{d^2\vec p_{cT}}
\frac{d^2N_{\bar{c}}}{d^2\vec p_{\bar{c}T}}W_\psi(\vec r, \vec k),
\label{CoalTrans}
\end{eqnarray}
under the assumption that the longitudinal momentum distributions
of (anti-)charm quarks are boost-invariant and they satisfy the
Bjorken correlation between spatial and momentum rapidities,
$\eta=y$. In Eq. (\ref{CoalTrans}) $\vec r$ and $\vec k$ are,
respectively, the distance and the relative momentum between two
charm quarks. We apply here Wigner functions, Eqs. (\ref{WigGau}),
(\ref{wigner10}), (\ref{wigner1s}), and (\ref{wigner2s}), and
carry out the integration over $\vec r$; {\allowdisplaybreaks
\begin{eqnarray}
&& \int d^3\vec r W_\psi(\vec r, \vec k) \nonumber \\
&& =\left\{ \begin{array}{ll} (2\sqrt{\pi}\sigma)^3
e^{-k^2\sigma^2} &
\psi_s^G; J/\psi  \\
\frac{2}{3}(2\sqrt{\pi}\sigma)^3 e^{-k^2\sigma^2}\sigma^2k^2 &
\psi_p^G; \chi_c  \\
\frac{2}{3}(2\sqrt{\pi}\sigma)^3 e^{-k^2\sigma^2}\Big(
\sigma^2k^2-\frac{3}{2}\Big)^2 & \psi_{10}^G; \psi(2S) \\
64\pi\frac{a_0^3}{(a_0^2k^2+1)^4} & \psi_{1S}^C; J/\psi \\
8\pi a_0^3\frac{(a_0^2k^2-1/4)^2}{(a_0^2k^2+1/4)^6} & \psi_{2S}^C;
\psi(2S) \end{array} \right. \label{WigIntdr}
\end{eqnarray}}
where the superscript $G$ and $C$ are denoted by, respectively,
the Gaussian and Coulomb hadron wave functions used in the Wigner
function.

As we clearly see in Eq. (\ref{WigIntdr}), we get surprisingly
simple results from such complicated Wigner functions, Eqs.
(\ref{wigner1s}) and (\ref{wigner2s}). We further notice that the
Wigner function for $s$-wave states becomes, after the analytic
integration over $\vec r$, the absolute value square of each wave
function in momentum representation; $|\tilde{\psi}_{s}(\vec
k)|^2=(2\sqrt{\pi}\sigma)^3 e^{-k^2\sigma^2}$,
$|\tilde{\psi}_{10}(\vec k)|^2=2/3(2\sqrt{\pi}\sigma)^3
e^{-k^2\sigma^2}(\sigma^2k^2-3/2)^2$, $|\tilde{\psi}_{1S}(\vec
k)|^2=64\pi a_0^3 /(a_0^2k^2+1)^4$, and $|\tilde{\psi}_{2S}(\vec
k)|^2=8\pi a_0^3 (a_0^2k^2-1/4)^2/(a_0^2k^2+1/4)^6$. From this we
deduce the general relation $\int d^3\vec r W_{\psi_l}(\vec r,
\vec k)=1/(2l+1)\sum_{-l}^l |\tilde{\psi}_{l}(\vec k)|^2$, which
is a new way of evaluating Wigner functions purely from hadron
wave functions without resorting to the definition of the Wigner
function, Eq. (\ref{wigner}) when the integration of the Wigner
function over space is necessary.

As has been discussed in Ref. \cite{Cho:2011ew}, we omit the
contribution from the longitudinal momentum in the Wigner function
at midrapidities, $y$=0; the relative momentum between charm
quarks becomes, $\vec k=(\vec p_{cT}'-\vec p_{\bar{c}T}')/2$ with
$\vec p_{cT}'$ and $\vec p_{\bar{c}T}'$ being the transverse
momenta in the charmonium frame \cite{Scheibl:1998tk, Oh:2009zj}.
We suppose that the hadronization volume $V$ is 1000 (2700) fm$^3$
for central Au+Au (Pb+Pb) collisions at $\sqrt{s_{NN}}$ =200 GeV
(2.76 TeV) at RHIC (LHC), respectively. The oscillator frequency
$\omega$ for the Gaussian Wigner function has been determined for
each charmonium from the relation between the mean square distance
and $\sigma$; $\langle
r^2\rangle_{J/\psi}$=3/2$\sigma_{J/\psi}^2$, $\langle
r^2\rangle_{\chi_c}$=5/2$\sigma_{\chi_c}^2$, and $\langle
r^2\rangle_{\psi(2S)}$=7/2$\sigma_{\psi(2S)}^2$. Using the quark
separation distances $r_0$=0.50, 0.72, and 0.90 fm
\cite{Satz:2005hx} we obtain oscillator frequencies 311.4, 250.3,
and 224.3 MeV, respectively, for $J/\psi$, $\chi_{c1}$, and
$\psi(2S)$. For the Coulomb Wigner functions $a_0$ has been
obtained from the relation between $\langle r^2\rangle$ and $a_0$
for both $J/\psi$ and $\psi(2S)$ mesons; $\langle
r^2\rangle=3a_0^2$ for $J/\psi$ and $42a_0^2$ for $\psi(2S)$.
Using the same $r_0$ we get $a_0^{J/\psi}$=0.289 fm, and
$a_0^{\psi(2S)}$=0.139 fm.

\begin{figure}[!t]
\begin{center}
\includegraphics[width=0.48\textwidth]{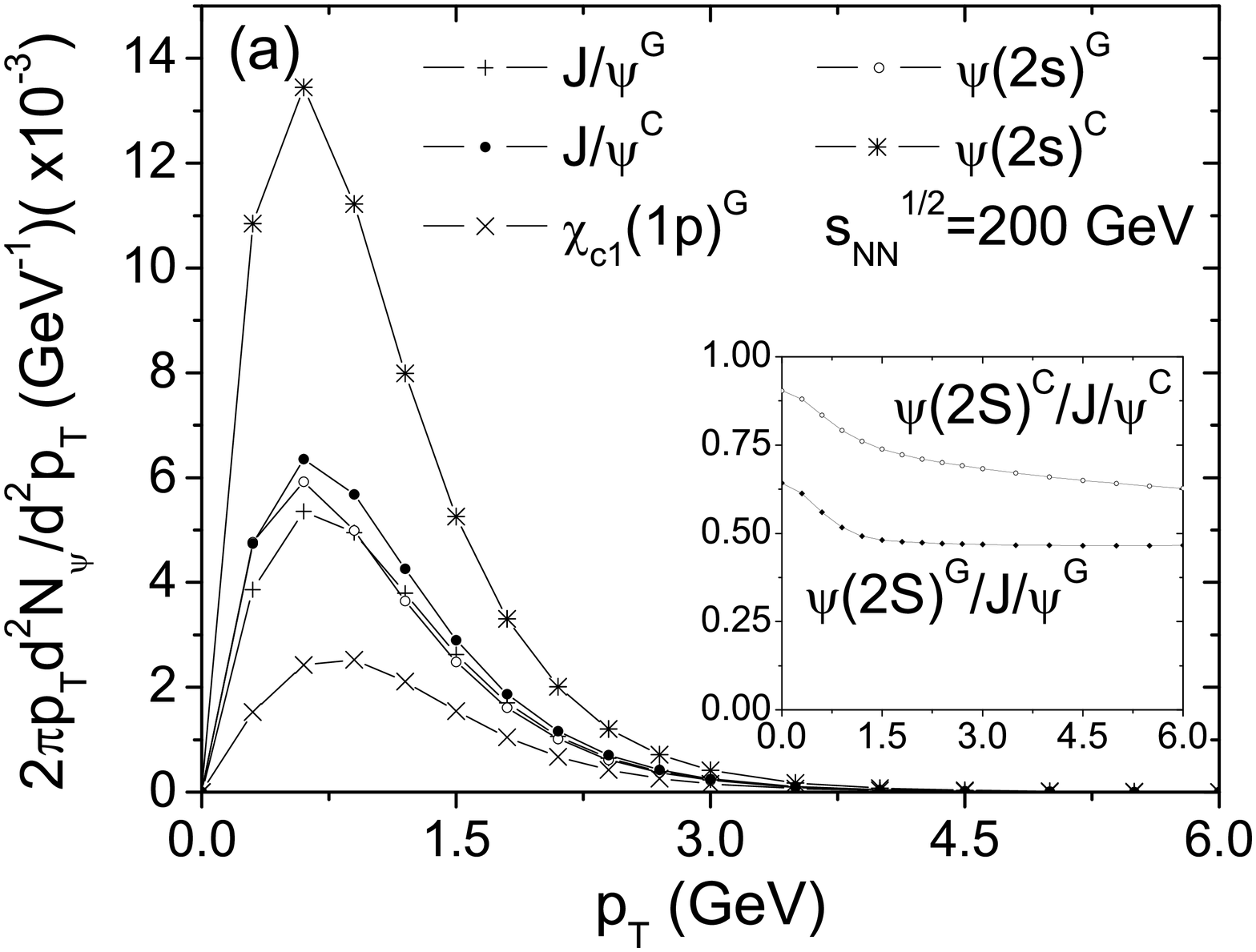}
\includegraphics[width=0.49\textwidth]{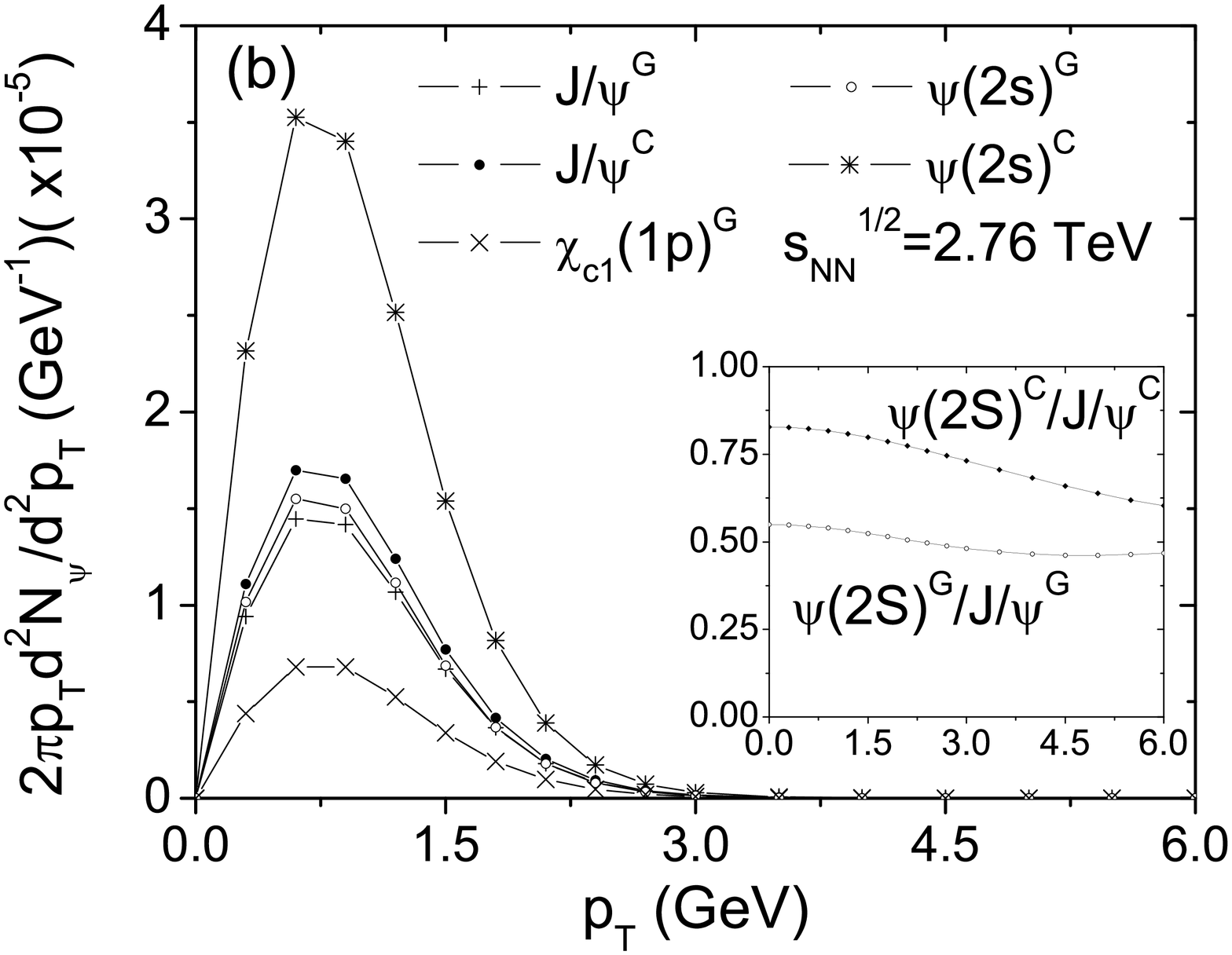}
\end{center}
\caption{$dN_{\psi}/dp_T$ of charmonia produced form charm and
anti-charm quarks by recombination for $\sqrt{s_{NN}}$=200 GeV at
RHIC (a) and for $\sqrt{s_{NN}}$=2.76 TeV at LHC (b). The ratio of
$p_T$ spectrum between $\psi(2S)$ and prompt $J/\psi$ mesons is
shown in the inset.} \label{pT_charmonia}
\end{figure}

We use the $p_T$ spectrum of charm quarks for RHIC
$\sqrt{s_{NN}}$=200 GeV, $d^2N_c/d^2\vec
p_T=19.2(1+p_T^2/36)/(1+p_T/3.7)^{12}/(1+e^{0.9-2p_T})
(0.8e^{-p_T/1.2}+0.6e^{-p_T/15})$ \cite{Oh:2009zj}, obtained from
heavy quark $p_T$ distribution in p+p collisions at the same
energy both scaled by number of binary collisions and multiplied
by the heavy quark energy loss estimation. Since there is no charm
quark $p_T$ distribution at hadronization available for LHC, we
apply the initial charm quark $p_T$ distribution, $dN_c/dp_T=
p_T/(1+0.379p_T^2)^{5.881}$ \cite{Lang:2013wya}, obtained from a
fit to PYTHIA evaluations for LHC $\sqrt{s_{NN}}$=2.76 TeV. The
$p_T$ is in unit of GeV at both distributions. Using these with
Eq. (\ref{WigIntdr}) we evaluate Eq. (\ref{CoalTrans}) produced by
recombination without feed-down contributions, and show the
results in Fig. \ref{pT_charmonia}.

We obtain the $p_T$ distribution for the $p$-wave state, the
$\chi_{c1}$ smaller than that of the $s$-wave state, the $J/\psi$.
On the other hand, we get the large $p_T$ distribution of the
excited state, the $\psi(2S)$, compared to that of the ground
state from both Gaussian and Coulomb Wigner functions. The
discrepancy is more significant when Coulomb Wigner functions have
been applied. We also show in the inset of Fig. \ref{pT_charmonia}
the ratio of the $p_T$ spectrum between the $\psi(2S)$ and
$J/\psi$ meson after considering prompt $J/\psi$ mesons,
$d^2N_{\psi(2S)^{C(G)}}/d^2\vec
p_T/d^2N_{J/\psi^{C(G)}}^{\mathrm{prompt}}/d^2\vec p_T$, by
assuming that the $p_T$ spectrum of the daughter $J/\psi$ meson is
same as that of the excited states after their decays;
$d^2N_{J/\psi^{C(G)}}^{\mathrm{prompt}}/d^2\vec
p_T$=$d^2N_{J/\psi^{C(G)}}/d^2\vec
p_T$+0.348~$d^2N_{\chi_{c1}^G}/d^2\vec p_T$ +0.198~
$d^2N_{\chi_{c2}^G}/d^2\vec p_T$+0.603~$
d^2N_{\psi(2S)^{C(G)}}/d^2\vec p_T$ \cite{Beringer:1900zz}. We see
that the $p_T$ distribution of the $\psi(2S)$ are comparable to
that of the $J/\psi$ at both RHIC and LHC.

The $p_T$ spectrum of the $\psi(2S)$ is expected to be smaller
than that of the $J/\psi$ since the $\psi(2S)$ is heavier than the
$J/\psi$ by about 600 MeV. We have found that the enhanced
$\psi(2S)$ production compared to the $J/\psi$ shown in Fig.
\ref{pT_charmonia} is attributable to the relative distribution of
their wave functions in momentum space. We remind that the wave
function of excited states is spread in space more broadly than
that of the ground state. In other words, the Coulomb wave
function of the $\psi(2S)$ is more localized around the origin in
momentum space than that of the $J/\psi$, Eq. (\ref{WigIntdr}),
leading to the larger $p_T$ distribution for $\psi(2S)$ mesons
compared to $J/\psi$ mesons when produced from charm quarks by
recombination. On the other hand, Gaussian wave functions in space
are again Gaussian in momentum space, and therefore distributions
in wave functions of excited states in space are preserved also in
momentum space, resulting in the smaller $p_T$ distribution for
the $p$-wave state $\chi_{c1}$. The Gaussian wave function of the
$\psi(2S)$ is spread further in momentum space, but the part of
its wave function localized between the origin and the node in
momentum space plays a significant role, contributing to the
similar $p_T$ distribution of the $\psi(2S)$ compared to that of
the $J/\psi$.

The real wave functions of charmonium states would be neither
Gaussian nor Coulomb wave functions. The harmonic oscillator
potential is strongly confining ($\propto r^2$) whereas the
Coulomb potential is loosely confining ($\propto r^{-1}$). Two
results shown in Fig. \ref{pT_charmonia}, therefore, provide
references to experimental measurements of the $p_T$ distribution
ratio between $J/\psi$ and $\psi(2S)$ mesons resulted from Coulomb
and linear potentials between charm quarks. We expect the study of
$p_T$ distributions of charmonium states to provide one way of
understanding the potential between charm quarks.

In summary, we have studied the $\psi(2S)$ meson production by
recombination of charm quarks in heavy ion collisions. We have
evaluated Wigner functions for the $\psi(2S)$ meson from both
Gaussian and Coulomb wave functions, and have investigated the
wave function dependence of the $p_T$ distribution. We have found
that the $p_T$ spectrum of $\psi(2S)$ mesons is similar or larger
compared to that of $J/\psi$ mesons due to their intrinsic wave
function differences in space. We suggest that the enhanced $p_T$
spectrum of the $\psi(2S)$ meson originated from its wave function
can present one way of understanding the increased ratio of the
$R_{AA}$ between $J/\psi$ and $\psi(2S)$ mesons with centrality,
recently measured at 3 $<p_T<$ 30 GeV in 1.6 $<|y|<$ 2.4 by CMS
Collaboration \cite{CMS}. We expect the precise measurements of
the same $R_{AA}$ ratio at midrapidity in the future to confirm
the $\psi(2S)$ production by recombination along with the
dependence of charmonia production on their wave functions.

\textit{Acknowledgements} We are grateful to Su Houng Lee for
fruitful discussions. This work was supported by the Korean
Ministry of Education through the BK21 PLUS program.

\end{document}